\documentclass[12pt]{article}

\newcommand{\bbr}{I\!\! R}
\newcommand{\bbz}{Z\!\!\! Z}

\newcommand{\2}{$^2$}
\newcommand{\3}{$^3$}
\newcommand{\4}{$_4$}

\newcommand{\x}{arXiv:}

\usepackage{graphicx}

\begin{document}
\thispagestyle{empty}
\begin{center}

\null \vskip-1truecm \vskip2truecm {\bf Answering a Basic
Objection to Bang/Crunch Holography
\\} \vskip1truecm Brett McInnes \vskip1truecm

 National University of Singapore

email: matmcinn@nus.edu.sg\\

\end{center}
\vskip1truecm \centerline{ABSTRACT} \baselineskip=15pt
\medskip
The current cosmic acceleration does \emph{not} imply that our
Universe is basically de Sitter-like: in the first part of this
work we argue that, by introducing matter into \emph{anti-de
Sitter} spacetime in a natural way, one may be able to account for
the acceleration just as well. However, this leads to a Big
Crunch, and the Euclidean versions of Bang/Crunch cosmologies have
[apparently] disconnected conformal boundaries. As Maldacena and
Maoz have recently stressed, this \emph{seems} to contradict the
holographic principle. In the second part we argue that this
``double boundary problem" is a matter not of geometry but rather
of how one chooses a conformal compactification: if one chooses to
compactify in an unorthodox way, then the appearance of
disconnectedness can be regarded as a \emph{coordinate effect}.
With the kind of matter we have introduced here, namely a
Euclidean axion, the underlying compact Euclidean manifold has an
unexpectedly non-trivial topology: it is in fact one of the 75
possible underlying manifolds of flat compact four-dimensional
Euclidean spaces.

\vskip3.5truecm
\begin{center}

\end{center}

\newpage

\addtocounter{section}{1}
\section*{1. de Sitter $\dots$ or Anti-de Sitter?}
From the fact \cite{kn:carroll}\cite{kn:riess} that our Universe
is accelerating, it is often deduced that our spacetime basically
resembles de Sitter spacetime. But is this justified?

There are in fact general reasons
\cite{kn:witten1}\cite{kn:susskind}\cite{kn:fischler} for thinking
that, despite appearances, our world is basically unlike de Sitter
spacetime. For example, \emph{holography} does not work very well
for de Sitter-like spacetimes. Clearly de Sitter spacetime itself
does not have a holographic Euclidean version, since its standard
Euclidean version has no boundary [see however \cite{kn:sim}], and
its Lorentzian holography
\cite{kn:strominger}\cite{kn:klemm1}\cite{kn:minic} leads to
problems \cite{kn:goheer} which have only been incompletely
resolved \cite{kn:mcinnesschwarz}\cite{kn:silver}\cite{kn:klemm2}.

It is therefore important to note that a Universe which
accelerates only temporarily --- and we have no reason to believe
that it does otherwise --- need not be related to de Sitter
spacetime at all: in fact, \emph{it can be modelled by a suitably
modified version of anti-de Sitter spacetime.} In view of all
this, it is reasonable to ask: can we dispense with de Sitter
spacetime altogether, and construct a complete cosmology using
only anti-de Sitter physics?

AdS\4 itself is of course completely unacceptable as a
cosmological model, since it does not expand, has no Big Bang, and
is not globally hyperbolic [so that it cannot be understood as a
FRW cosmology --- it has a FRW coordinate system, but this only
covers a small part of the spacetime]. All of these blemishes are
removed, however, by introducing matter into AdS\4. In fact, by
introducing scalar matter into AdS\4 \cite{kn:cardenas} one can
produce a globally hyperbolic spacetime which does have a Bang and
also a \emph{temporary} burst of acceleration. [See also
\cite{kn:decay}.] Such spacetimes are therefore ideally suited to
answering questions about the global structure of spacetimes which
accelerate and yet \emph{are not asymptotically de Sitter,}
because the acceleration is only temporary. [For another approach
to using AdS in cosmology, see \cite{kn:cvetic}.]

The cosmological models obtained in this way underline a
fundamental fact which deserves more attention than it has
received: AdS\4 is unstable to singularity formation. The
introduction of \emph{any} amount of \emph{any} form of matter
[other than vacuum energy itself] satisfying the Strong Energy
Condition causes singularities to develop, producing a spacetime
\emph{which is not a small perturbation} of AdS\4. [This follows
from the Hawking-Penrose cosmological singularity theorem
\cite{kn:hawking}. See also \cite{kn:leblond} for a good recent
discussion of this theorem. Note that singularities will
\emph{usually} form even if the matter does not satisfy the SEC.]
Singular versions of AdS\4 are therefore the generic spacetimes
with negative cosmological constants, and so it is quite natural
for an AdS\4-derived spacetime to accommodate a Big Bang [and to
be globally hyperbolic].

However, putting a Bang into AdS\4 has a price: it almost
inevitably produces a Crunch. This is because, as the Universe
expands, it dilutes all forms of matter and radiation [except
phantom matter \cite{kn:caldwell}, which we do not consider here]
more rapidly than a cosmological constant, so the latter will
become increasingly important as the expansion proceeds. The
negative cosmological constant of AdS\4 will inevitably halt the
expansion and cause a contraction. If the Universe is still
contracting when matter and radiation again become dominant, there
will be a Crunch at the end just as there was a Bang at the
beginning. Hence a quasi-realistic cosmological model arising from
AdS\4 has \emph{both} a Crunch and a Bang. [For the basic theory
of Crunch cosmologies, see \cite{kn:linde}.]

Because these AdS-derived cosmologies are singular, their global
structure is radically different to that of AdS itself, even
though their matter content may be a small perturbation away from
the ``matter content" [the negative cosmological constant] of AdS.
In particular, there is no longer any timelike spatial infinity.
Hence it is no trivial matter to extend the AdS holographic
description \cite{kn:maldacena} to them, and indeed one might
think that this is not possible: perhaps holography forbids a
Crunch. Recently, however, this question has been investigated
\cite{kn:horowitz} in the ``intermediate" case where singularities
are not forbidden but the boundary conditions are forced to be
asymptotically AdS. The conclusion is that reasonable
asymptotically AdS data do evolve to a Crunch, and that this may
well be correctly represented in the CFT at infinity. Because the
spacetimes considered in \cite{kn:horowitz} have boundary
conditions specified at spatial infinity, they are not
cosmological spacetimes, but this work does show that holography,
and in particular string/M theory, is \emph{not} fundamentally
incompatible with a Crunch. This prompts the question: how is it
possible to obtain a holographic description of a Crunch in the
context of a genuine cosmological model?

Maldacena and Maoz \cite{kn:maoz} have recently argued that it
might be possible to establish such a description of AdS-derived
Bang/Crunch cosmologies by focusing primarily on the
 \emph{Euclidean} domain. In fact, the Euclidean
versions of Bang/Crunch cosmologies are automatically non-singular
and otherwise [apparently] very well-behaved. Thus while, for
example, a Lorentzian Crunch cannot be associated with a critical
point of a superpotential, it may well be possible to achieve such
a description of the non-singular object which replaces the Bang
and the Crunch, that is, the Euclidean conformal boundary.
\emph{The Maldacena-Maoz proposal is that singular Bang/Crunch
cosmological spacetimes are described by a dual field theory
defined on the non-singular boundary of the Euclidean version.}
Thus, the Euclidean version of a Bang/Crunch cosmology is taken to
be the fundamental version, the properties of the Lorentzian
spacetime being derived from it.

Now the Euclidean manifolds considered by Maldacena and Maoz
apparently have conformal compactification with boundaries
\emph{having two disconnected components}. Thus, the problem of
obtaining a holographic understanding of cosmology seems to
involve extending the [Euclidean] AdS/CFT correspondence to spaces
with multiple boundaries. This raises two issues, one
mathematical, the other physical.

The \emph{mathematical} issue is that it is not always
\emph{possible} to have two boundaries when physically reasonable
conditions are imposed on the geometries of the bulk and the
boundary, a point first made in \cite{kn:yau}. The subtle way in
which the Maldacena-Maoz cosmologies [including, especially,
versions with temporary acceleration] evade the Witten-Yau theorem
was the subject of \cite{kn:mcinnes}, and we shall not consider
this question further here: we merely note that it is \emph{not}
in fact possible to prohibit multiple boundaries in this way.

The \emph{physical} issue is that having two [or more] boundary
components for one bulk seems to be a blatant contradiction
\cite{kn:yau} of the holographic philosophy, which is founded on a
putative one-to-one bulk/boundary correspondence. \emph{In
general} this is a very deep question: some relevant ideas are
advanced in \cite{kn:maoz} [see also \cite{kn:maoz2}] and, in the
lower-dimensional context, in \cite{kn:balasub} and
\cite{kn:gukov}. We wish to argue, however, that in the
\emph{cosmological} context [though not in general] this is
something of a red herring. What is special about the cosmological
context is that the two boundary components have to have the
\emph{same topology}, since topology ``change" in the Euclidean
version would entail topology change in the course of the
evolution of the Lorentzian version, and it is well known that
this leads to various difficulties
\cite{kn:geroch}\cite{kn:tipler} under physically reasonable
conditions. This means that we should ask whether the ``two"
boundary components \emph{might perhaps be one and the same}.
Obviously this would solve the problem in the simplest and most
direct manner.

We shall present a concrete example of a quasi-realistic anti-de
Sitter based cosmology. We begin, motivated by the remarks in
\cite{kn:susskind}, by formulating a Euclidean superpotential for
an axion field, and show briefly how to adjust the parameters by
using observational data. We then show that there is a way of
interpreting the Euclidean metric such that conformal infinity is
clearly connected, despite the existence of both a Bang and a
Crunch in the Lorentzian spacetime. If we accept this [mildly]
unorthodox way of assigning a conformal infinity to a non-compact
Euclidean manifold, then the double boundary problem simply arises
from a bad choice of coordinates; for the conformal
compactification is \emph{not} a manifold-with-boundary but rather
simply a compact manifold, one of the possible underlying
manifolds of flat compact spaces in four dimensions.

It turns out that there are several possible topological
structures for this manifold, just as, in the more familiar
two-dimensional case, a flat compact manifold can be either a
torus or a Klein bottle. We argue that the choice should be made
on physical grounds, taking into account the fact that the matter
here is axionic. The result is that the Euclidean conformal
compactification is topologically non-trivial. Thus the real
significance of the double boundary problem is revealed: it is
simply telling us that the Euclidean version has a more intricate
topology than at first appears. \emph{This structure is, however,
fully compatible with the holographic principle.}

\addtocounter{section}{1}
\section*{2. A Euclidean Axion}
As we have explained, the Euclidean spacetimes in which we are
interested are topologically non-trivial in the sense that [in the
most obvious interpretation of the metric] the boundary is
disconnected; Maldacena and Maoz describe them as [Euclidean]
``wormholes", though of course the Lorentzian interpretation is
not a wormhole in the usual sense. Thus the matter we shall
introduce into Euclidean AdS\4 has to be carefully chosen; most
forms of matter do not lead to topologically non-trivial spaces.
Maldacena and Maoz \cite{kn:maoz} use meron and instanton
Yang-Mills configurations, which automatically satisfy the Strong
Energy Condition and hence cannot lead to even temporary cosmic
acceleration. In order to improve on this we need a kind of matter
which can lead to

\medskip
[a] temporary acceleration in the Lorentzian version and

\medskip
[b] topological non-triviality in the Euclidean version.
\medskip

In view of point [a], the natural kind of matter to consider is
scalar [or perhaps pseudoscalar] matter, since it is well known
\cite{kn:ratra} that such matter [``quintessence"] can lead to
acceleration. However, it is also known that scalar fields cannot
lead to topologically non-trivial asymptotically flat Euclidean
configurations, as in point [b], unless the field itself is
complexified upon passing between the Lorentzian and Euclidean
domains. For this the reader may consult \cite{kn:GS}; see also
the comments in \cite{kn:wald} and \cite{kn:coule}. All these
works are directly relevant only to the asymptotically \emph{flat}
case; but it can be shown that the same requirement for
complexification holds in the asymptotically anti-de Sitter case
also. [There are technical complications: it is no longer true, as
in the asymptotically flat case, that topologically non-trivial
configurations are only possible when the Euclidean scalar field
makes a negative contribution to the Ricci curvature; however,
there are other conditions, replacing this one, which still
require field complexification. See \cite{kn:mcinnes} for the
details.]

Complexification of the field arises most naturally for
\emph{axions}. Thus points [a] and [b] above lead us naturally to
the assumption that the Euclidean version of the Universe is
basically like Euclidean AdS\4, but differs from it because of the
presence of a Euclidean axion, that is, a pseudo-scalar with a
periodic potential. [For the use of \emph{Lorentzian} axions in
cosmology, see for example
\cite{kn:choi}\cite{kn:kim}\cite{kn:gonzalez}.]

In \cite{kn:susskind}, Hellerman et al show that permanently
accelerating quintessence spacetimes do not evolve towards a
supersymmetric
 state described by a critical point of a superpotential, and this is adduced
  as further evidence for the incompatibility of de Sitter-like spacetimes
  with string theory. In anti-de Sitter cosmology, where
  the universe ends in a Crunch, we have even less reason to
  believe that the system is evolving towards a supersymmetric
  state. However, in the context of the Maldacena-Maoz approach,
  in which the Euclidean version is fundamental, we can work towards
  answering this objection by
  ensuring that the asymptotically AdS regions of the \emph{Euclidean}
  version do correspond to critical points of a superpotential. Since
  we are dealing with a Euclidean axion, which should have a
  periodic potential, we can express the superpotential W$^+$($\varphi^+$),
  for a single Euclidean field $\varphi^+$, as a
  Fourier series. [To avoid confusion, a
superscript + will
 be used where necessary to emphasise that a quantity is ``Euclidean", with the Lorentzian version
  being denoted by a minus sign.] We then have
\begin{equation}\label{eq:ABRACADABRA1}
\textup{W}^+(\varphi^+)\; =\; \sum_{k=1}^\infty\;\textup{C}_k \,
\textup{sin}(k\,\sqrt{{{4\pi}\over{\varpi}}}\;\;\varphi^+)\;+\;{{1}\over{2}}\,\textup{B}_0\;+\;\sum_{k=1}^\infty\;\textup{B}_k
\, \textup{cos}(k\,\sqrt{{{4\pi}\over{\varpi}}}\;\;\varphi^+).
\end{equation}
Here $\varpi$ is a  positive constant; the way in which it appears
is for later convenience.

Since the sign of a pseudoscalar field has a physical
significance, it is particularly interesting to consider
\emph{odd} superpotentials, so we restrict to
\begin{equation}\label{eq:ABRACADABRA2}
\textup{W}^+(\varphi^+)\; =\; \sum_{k=1}^\infty\;\textup{C}_k \,
\textup{sin}(k\,\sqrt{{{4\pi}\over{\varpi}}}\;\;\varphi^+).
\end{equation}
It will turn out that the higher frequency modes of this expansion
are important only extremely near to the Bang and the Crunch. For
simplicity we postpone their consideration and consider [for now]
only a single term in this expansion, namely the first:
\begin{equation}\label{eq:A}
\textup{W}^+(\varphi^+)\; =\; \textup{C\,
sin}(\sqrt{{{4\pi}\over{\varpi}}}\;\;\varphi^+)\;;
\end{equation}
The effect of replacing this superpotential by the k-th
higher-order term is of course obtained simply by replacing
$\varpi$ by $\varpi$/k$^2$.

Because the axion is complexified when passing between the
Euclidean and Lorentzian domains, the relation between potential
and superpotential here
 is given [in four dimensions] by
\begin{equation}\label{eq:B}
\textup{V}^+(\varphi^+)\; =\; -\;8{(\textup{W}^+)^{\prime}}^{\;2}
\; - \; 96\pi (\textup{W}^{+})^2.
\end{equation}
For pure Euclidean AdS\4 with all sectional curvatures equal to
$-$1/L\2, obtained when (W$^+$)$^\prime$ vanishes, the potential
should have its usual value for a negative cosmological constant,
that is, $-$3/(8$\pi$L$^2$). This fixes the constant C in
(\ref{eq:A}) at the value 1/(16$\pi$L), so we in fact have
\begin{equation}\label{eq:C}
\textup{W}^+(\varphi^+)\; =\; {{1}\over{16\pi \textup{L}}}\;
\textup{sin}(\sqrt{{{4\pi}\over{\varpi}}}\;\;\varphi^+).
\end{equation}
Substituting this into (\ref{eq:B}) we obtain, after some simple
manipulations,
\begin{equation}\label{eq:D}
\textup{V}^+(\varphi^+)\; =\; - {{3}\over{8\pi
\textup{L}^2}}\;+\;\textup{V}^+_{\textup{Axion}},
\end{equation}
where
\begin{equation}\label{eq:E}
\textup{V}^+_{\textup{Axion}}\;=\;{{3 \;-\;\varpi^{-1}}\over{8\pi
\textup{L}^2}}\;\textup{cos}^2(\sqrt{{{4\pi}\over{\varpi}}}\;\;\varphi^+).
\end{equation}
Clearly this system \emph{appears} to be Euclidean AdS\4, with
``energy" density $-$3/(8$\pi$L$^2$), into which we have
introduced a matter field with a potential V$^+_{\textup{Axion}}$.
Notice however that the AdS\4 energy density does not really arise
independently of the energy density of the axionic matter being
introduced
--- both have a common origin in the original superpotential.

The matter potential V$^+_{\textup{Axion}}$ will be positive
provided that
\begin{equation}\label{eq:F}
\varpi\;\ > \;{{1}\over{3}}\;;
\end{equation}
this will always be satisfied in the sequel since it will turn out
that the physically interesting values of $\varpi$ are larger than
unity. Thus the matter we are ``introducing" into AdS\4 is not
very exotic.

The combined Einstein-axion field equations for this potential can
be solved exactly if we assume that the geometry takes the
Euclidean FRW form with flat ``spatial" sections. To be specific,
we assume that the sections are compact and flat; in the simplest
case they are tori, so that the metric takes the form
\begin{equation}\label{eq:I}
g^+\; = \;(dt^+)^2 \;+\; A^2\;a(t^+)^2[d\theta_1^2\; +
\;d\theta_2^2 \;+\; d\theta_3^2],
\end{equation}
where A measures the circumferences of the torus when a(t$^+$),
the scale factor, is equal to unity.

There are several motivations for taking the ``spatial" sections
to be compact. In the generalized [Euclidean] AdS/CFT
correspondence, it is best if the CFT is defined on a compact
space, so as to avoid problems with non-unique correlation
functions --- see Section 2.3 of \cite{kn:witten2}. Note that,
even in the Lorentzian version of AdS/CFT, the CFT is really
defined not on Minkowski space but rather
\cite{kn:witten2}\cite{kn:orbifold} on its conformal \emph{
compactification} or on the latter's universal cover, both of
which have compact spatial sections of topology S$^3$. We take all
this as strong evidence that cosmological holography works best
--- or perhaps only --- for cosmologies with compact ``spatial"
sections in both the Euclidean and Lorentzian versions. Again,
there are in fact very interesting string cosmological models in
which it is essential that the sections should be both flat and
compact: see \cite{kn:brand}\cite{kn:watson} and also, for a quite
different model, \cite{kn:paban}.

Another motivation for taking the ``spatial" sections to be
compact is the following. The observational data \cite{kn:riess}
suggest that the spatial sections of our Universe are close to
being flat. [The ``spatial" sections of the Euclidean version of
the manifold have the same geometry and topology as the spatial
sections of the Lorentzian version.] Naturally we should be aware
that the data may be misleading us: they are equally compatible
with a small positive curvature. However, it is interesting to
consider the possibility that the data are hinting that the
spatial sections of our world \emph{really are} flat, by which we
mean that they could have a topological structure which
\emph{forbids} them to have completely positive or completely
negative curvatures. This is a property \cite{kn:lawson} shared by
all of the ten different \emph{compact} three-manifolds
\cite{kn:wolf} which can be flat, but not by $\bbr^3$ [which is
the underlying topology of both flat and hyperbolic space in the
simply connected cases]. Thus the data pointing to ``exact"
flatness suggest that the spatial sections are modelled on one of
these ten manifolds. We have chosen the torus T\3 only for reasons
of simplicity: the other nine possibilities will not be discussed
here.

The combined Euclidean Einstein-matter equations for the potential
given in (\ref{eq:D}) and (\ref{eq:E}) yield
\begin{equation}\label{eq:J}
(\dot{\varphi^+})^2\;=\;{{1}\over{4\pi \varpi
\textup{L}^2}}\;\textup{cos}^2(\sqrt{{{4\pi}\over{\varpi}}}\;\;\varphi^+).
\end{equation}
This provides us with a useful consistency check: since the right
hand side of this equation obviously does not change sign whether
or not $\varphi^+$ is complexified when passing to the Lorentzian
domain, this equation only makes sense if $\varphi^+$ \emph{is}
complexified [since otherwise the left side would change sign when
t is complexified and the right side would not]. Thus indeed
$\varphi^+$ must be complexified, in the manner of an axion, when
transforming to the Lorentzian version.

The solution of (\ref{eq:J}) is given by
\begin{equation}\label{eq:K}
\varphi^+\;=\;\pm\sqrt{{{\varpi}\over{4\pi}}}\;\textup{cos}^{-1}(\textup{sech}({{t^+}\over{\varpi\textup{L}}})),
\end{equation}
where t$^+$ runs from $-\infty$ to $+\infty$, where the sign
agrees with that of t$^+$, and where cos$^{-1}$ is defined to take
its values between 0 and $\pi/2$. This solution will be important
below.

It is shown in \cite{kn:mcinnes} that the Einstein equation can be
solved exactly here: the solution for the metric takes the
surprisingly simple form
\begin{equation}\label{eq:M}
g^+(\varpi,A) = (dt^+)^2\; + \;
A^2\;\textup{cosh}^{2\varpi}({{t^+}\over {\varpi
\textup{L}}})\;[\,d\theta_1^2 \;+\; d\theta_2^2\; +
\;d\theta_3^2\,].
\end{equation}
This is [as foreseen] completely non-singular, with asymptotically
hyperbolic regions near t$^+$ = $\pm\infty$; that is, the metric
increasingly resembles that of hyperbolic space, the Euclidean
version of AdS\4. This can be seen explicitly, since this metric
has the approximate asymptotic form
\begin{equation}\label{eq:MM}
g^+(\varpi,A)\; \approx \; (dt^+)^2\; + \;
4^{-\varpi}A^2\;\textup{exp}(2|\,t^+|/\textup{L})\;[\,d\theta_1^2
\;+\; d\theta_2^2\; + \;d\theta_3^2\,],
\end{equation}
which is a well-known local form of the standard metric of
constant curvature $-$1/L\2. We see that the length scale here is
just L, the curvature scale set by the asymptotic cosmological
constant $-$3/(8$\pi$L$^2$). [There is no other natural length
scale, because there is none associated with the flat ``spatial"
directions, and the range of t is infinite.]

The metric $g^+(\varpi,A)$ in equation (\ref{eq:M}) can be
regarded, in the usual way, as a metric on the interior of a
manifold-with-boundary, the boundary consisting of \emph{two}
copies of some compact flat three-dimensional manifold [such as a
flat torus]. From equations (\ref{eq:C}) and (\ref{eq:K}) we see
that these flat three-manifolds correspond to the critical points
of the superpotential. If Euclidean holography is to make sense at
all in Bang/Crunch cosmology, it must apply to some kind of
Euclidean field theory defined on these flat spaces. A priori
there seems to be no reason why this should not work. Before
discussing this further, we turn to the question of the
plausibility of the Lorentzian version as a cosmological model.

\addtocounter{section}{1}
\section*{3. Cosmic Acceleration in Anti-de Sitter Physics}

The Lorentzian version of this space cannot be expected to share
any of the agreeable properties of $g^+(\varpi,A)$: the boundary
will be singular, there will be no superpotential with critical
points, and so on. The metric is of course
\begin{equation}\label{eq:L}
g^-(\varpi,A) = -\;(dt^-)^2 +
A^2\;\textup{cos}^{2\varpi}({{t^-}\over {\varpi
\textup{L}}})\;[d\theta_1^2 + d\theta_2^2 + d\theta_3^2].
\end{equation}
It can be shown that this is, as it appears to be, singular: there
is a Big Bang at t$^-$ = $-\pi\varpi$L/2, and a Big Crunch at
$+\pi\varpi$L/2. By complexifying $\varphi^+$ in the Euclidean
potential given by equations (\ref{eq:D}) and (\ref{eq:E}), we
obtain a Lorentzian potential which reveals that the Lorentzian
field $\varphi^-$ is a kind of quintessence:
\begin{equation}\label{eq:DUNG}
\textup{V}^-(\varphi^-)\; =\; - {{3}\over{8\pi
\textup{L}^2}}\;+\;\textup{V}^-_{\textup{Quintessence}},
\end{equation}
where
\begin{equation}\label{eq:H}
\textup{V}^-_{\textup{Quintessence}}\;=\;{{3
\;-\;\varpi^{-1}}\over{8\pi
\textup{L}^2}}\;\textup{cosh}^2(\sqrt{{{4\pi}\over{\varpi}}}\;\;\varphi^-).
\end{equation}
Such potentials have been extensively explored: similar ones were
investigated recently in \cite{kn:olive} and [in the context of
``racetrack inflation"] in \cite{kn:cline}. Of course, the model
we are considering here differs from these because, in equation
(\ref{eq:DUNG}), we are superimposing the ``quintessence"
potential on the negative ``potential" [cosmological constant] of
AdS\4. [For other exponential-like quintessence potentials see
\cite{kn:ratra}\cite{kn:townsend}\cite{kn:ish1}\cite{kn:kehagias}\cite{kn:jarv};
for other metrics similar to $g^-(\varpi,A)$, see
\cite{kn:erosh}.]

It is also interesting to compare this model with the one assumed
in the ``cyclic" scenario \cite{kn:turok2}. As in that case, the
total potential here can be negative; in our case, this is a
simple consequence of having a negative cosmological constant in
the background. In both cases we have a Bang and a Crunch.
However, there is an important dissimilarity also: we shall see
that, in our case, the negative potential [which reveals itself
through an equation-of-state parameter exceeding unity] is only
important near to the transition from expansion to contraction,
that is, well away from the Bang and the Crunch. In the cyclic
scenario, by contrast, it is very important that the
equation-of-state parameter should exceed unity \emph{near} to the
Bang and to the Crunch, because this is the way in which the
cyclic model avoids having large Belinsky-Khalatnikov-Lifschitz
anisotropies which might propagate through the ``bounce"
\cite{kn:turok1}. In our model there is no bounce. Nevertheless
our studies of cosmic holography will lead us to a geometric
structure which may be highly relevant to the ``cyclic" model, as
we shall discuss below.

One obtains $\varphi^-$ in the Lorentzian case either by solving
the field equation or directly by complexifying $\varphi^+$: the
result in either case is given by
\begin{equation}\label{eq:U}
\varphi^-\;=\;\pm\sqrt{{{\varpi}\over{4\pi}}}\;\textup{cosh}^{-1}(\textup{sec}({{t^-}\over{\varpi\textup{L}}})),
\end{equation}
where the sign follows that of t$^-$. Unlike $\varphi^+$,
$\varphi^-$ diverges towards the [singular] boundaries of the
Lorentzian spacetime, as one would expect.

The energy density of $\varphi^-$ can be computed in terms of the
scale function; the result \cite{kn:mcinnes} is
\begin{equation}\label{eq:EXTRA}
\rho\,(\varphi^-)\;=\;{{3}\over{8\pi\textup{L}^2}}\;a^{-2/\varpi}.
\end{equation}
The total energy density is the sum of this and the energy density
of the background AdS$_4$. Notice that if we had taken the k-th
order term in (\ref{eq:ABRACADABRA2}) instead of the first, then
the density of $\varphi^-$ would vary as $a^{-2\,k^2/\varpi}$. If
t$_r$ is some early time, say during the era of radiation
dominance [when all components of dark energy were of negligible
importance], and $\rho_r$ and $a_r$ are the values of the density
and the scale function then, we have in the general case
\begin{equation}\label{eq:MORE}
\rho\,(\varphi^-)/\rho_r\;=\;(a/a_r)^{-2\,k^2/\varpi}.
\end{equation}
Clearly the higher terms in the Fourier expansion
(\ref{eq:ABRACADABRA2}) make a contribution which dies away more
rapidly than the lower terms as the Universe expands, justifying
our emphasis on the first term. [If such terms were included, they
would be important at extremely \emph{early} times, but it would
not be consistent to do this since we are not including ordinary
matter and radiation in our model.]
\begin{figure}[!h]
\centering
\includegraphics[width=0.9\textwidth]{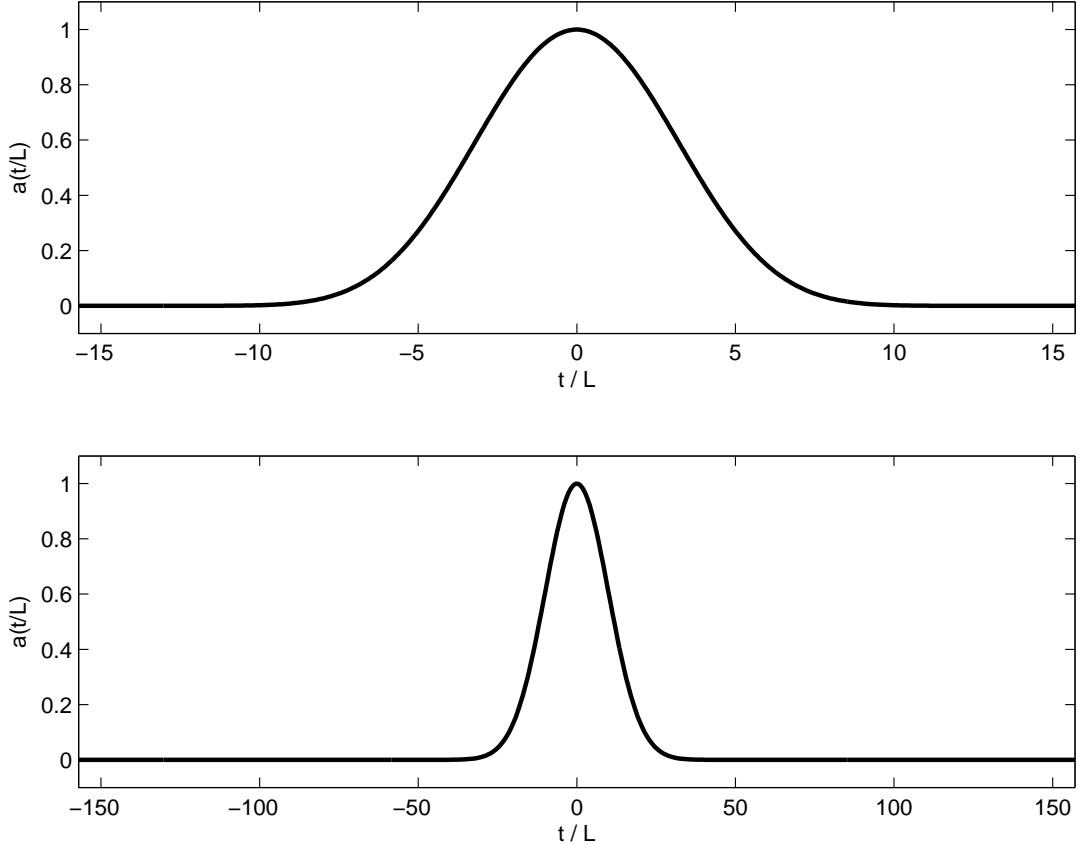}
\caption{Scale Factor for $\varpi$ = 10 [above] and 100 [below]}
\end{figure}

Graphs of the Lorentzian scale function, for two indicated values
of $\varpi$, are shown in Figure 1. It is clear from a glance at
these diagrams that \emph{these cosmologies can accelerate}; they
do so at times corresponding to those parts of the graphs which
are ``concave up". Note that t$^-$ = 0 is the moment of time
symmetry, beyond which the Universe begins to contract: so that
the present time corresponds to some \emph{negative} value of
t$^-$.

The natural length scale here is not set by the asymptotic
curvature, as it is in the Euclidean case. Instead, the total
duration of the spacetime, $\pi\varpi$L, sets the scale of this
system. Notice that this length scale need \emph{not} be the same
as that of the Euclidean version, which, as we saw, is simply
given by L. Thus the constant $\varpi$ measures the ratio of the
Lorentzian to the Euclidean length scales. In particular we can
consider the possibility that L is quite small; this is compatible
with a large observed Universe provided that $\varpi$ is large.

In the Euclidean version, the essential length scale reveals
itself in the AdS-like asymptotic regions [see equation
(\ref{eq:MM})], but this scale is concealed in the Lorentzian
version by the fact that there are no asymptotic regions in this
sense --- the spacetime is cut off by the Bang and the Crunch.
This is of interest because the values for L associated with
holographic theories can be small by cosmological standards. For
example, in an AdS/CFT scheme, one expects L to be related to the
coupling in the dual theory; large L will correspond to strong
coupling, small L to weak. Again, it has recently been emphasised
\cite{kn:bobby} that in many string compactifications, and
specifically in the Freund-Rubin case \cite{kn:denef}, the
requirement that the Kaluza-Klein modes be sufficiently massive as
to escape observation apparently requires the fundamental length
scale of four-dimensional spacetime to be far smaller than what we
observe. This discrepancy may be accounted for \cite{kn:bobby} by
quantum corrections, by supersymmetry breaking classical
corrections, and so on; in any case, there must be some parameter,
computable from these effects, which removes the discrepancy
between the fundamental and the observed length scales. We can
regard $\varpi$ as a simple concrete example of such a parameter.

As we shall see, observational data do in fact require $\varpi$ to
be fairly large [for any L], and so it is natural [though not
compulsory] to assume that L is small by cosmological standards;
so it is possible that one can work towards resolving these issues
in the way being advocated here: that is, the length scale of
four-dimensional space \emph{is} small, but only in the Euclidean
version. This would presumably require that the coupling in the
putative dual theory is weak.

The equation-of-state parameter w which gives the ratio of
pressure to density is \emph{not} a constant, being given instead
by
\begin{equation}\label{eq:N}
\textup{w} \;= \; -\;1\;+\;{{2}\over
{3\varpi}}\;\textup{cosec}^2({{t}\over {\varpi \textup{L}}});
\end{equation}
this is bounded below by [and comes arbitrarily close to] its
value ``at" the Bang,
\begin{equation}\label{eq:O}
\textup{w}_{\textup{Bang}} \;= \; -\;1\;+\;{{2}\over {3\varpi}}.
\end{equation}
As is well known, acceleration corresponds to w $< -$1/3, so
\emph{temporary acceleration will occur in this model} provided
that $\varpi$ exceeds unity, which is of course consistent with
the above discussion. In that case, if we define a time t$^*$ by
\begin{equation}\label{eq:P}
\textup{t}^* \; = \; -\;\varpi
\textup{L}\;\textup{sin}^{-1}\sqrt{{{1}\over {\varpi}}}\;,
\end{equation}
then the spacetime accelerates from the Bang until t$^-$ = t$^*$,
it then decelerates from t$^-$ = t$^*$ until t$^-$ = $|\,$t$^*|$
[having started to contract at t = 0], and then accelerates [while
contracting] from t$^-$ = $|\,$t$^*|$ until the Crunch. The
present time corresponds to some negative value of t$^-$ less than
t$^-$ = t$^*$. Notice that equation (\ref{eq:P}) shows that, if
$\varpi$ is large, $|\,$t$^*|$ is given approximately by
$\;\sqrt{\varpi}$ L, which is small as a fraction of the total
lifetime $\pi\varpi$L. Thus the interval of deceleration is
relatively short in this case [see Figure 1].

We observe from  that, unlike some pure quintessence spacetimes in
which w tends to $-$1, the w function here is bounded away from
$-1$. To reconcile this with the observations, which show that w
is currently rather close to $-$1, $\varpi$ has to be large. To be
more precise, w is currently \cite{kn:riess} no larger than about
$-$ 0.8 [and it may be far smaller], and since the cosec function
is bounded below by unity, we see from (\ref{eq:N}) that $\varpi$
has to be at least 3 or 4; in fact this is a gross underestimate.
Notice that this argument does \emph{not} depend on any
assumptions about the value of L; that is, $\varpi$ has to be
large [that is, significantly larger than it needs to be in order
to allow acceleration] in any case. Let us be a little more
precise about this.

If the present time is t$_0$
--- recall that this is a negative number
--- then define $\alpha$ to be the ratio of $|\,\textup{t}_0|$ to
the cosmic length scale $\varpi$L. Then $\alpha$ is an angle lying
between 0 and $\pi$/2. The values of the Hubble constant and of w
at the present time can be fairly strongly constrained by
observations. Much less stringent but nevertheless non-trivial
constraints can be placed \cite{kn:riess}\cite{kn:khoury} on the
rate at which w changes as we look back into cosmic history. [See
however \cite{kn:bass}.] Following
\cite{kn:khoury}\cite{kn:linder}, we measure this by means of the
parameter
\begin{equation}\label{eq:QQQ}
\textup{w}_a
\;=\;-\;a(\textup{t}_0){{\textup{dw}}\over{\textup{d}a}}(a(\textup{t}_0)),
\end{equation}
so that negative w$_a$ means that w is currently increasing. [This
parametrization is a significant improvement over the one used in
\cite{kn:mcinnes}, because it does not refer to the age of the
Universe, a datum which cannot be coherently included in this
model since it is so strongly affected by the matter content of
the early Universe.]

The observed quantities H$_0$, w$_0$, and w$_a$ are given in this
cosmology by
\begin{equation}\label{eq:QQ}
\textup{H}_0\;=\;{{1}\over{\textup{L}}}\;\textup{tan}(\alpha),
\end{equation}
\begin{equation}\label{eq:RR}
\textup{w}_0 \;= \; -\;1\;+\;{{2}\over
{3\varpi}}\;\textup{cosec}^2(\alpha),
\end{equation}
\begin{equation}\label{eq:SS}
\textup{w}_a
\;=\;-\;{{4\;\textup{cos}^2(\alpha)}\over{3\varpi^2\;\textup{sin}^4(\alpha)}}.
\end{equation}
Notice that the theory predicts that w must be increasing at the
present time.

Solving these three equations for the three unknowns L, $\varpi$,
and $\alpha$, we have
\begin{equation}\label{eq:QQA}
\textup{L}\;=\;{{\sqrt{\kappa^{-1}\;-\;1}}\over{\textup{H}_0}},
\end{equation}
\begin{equation}\label{eq:RRA}
\varpi \;= \; {{2}\over {3\,[1\;+\;\textup{w}_0][1\;-\;\kappa]}},
\end{equation}
\begin{equation}\label{eq:SSA}
\alpha\;=\;\textup{cos}^{-1}\sqrt{\kappa},
\end{equation}
where
\begin{equation}\label{eq:SSAA}
\kappa\;=\;{{|\,\textup{w}_a|}\over{3\,[1\;+\;\textup{w}_0]^2}}.
\end{equation}
The current value of the scale function is given by
\begin{equation}\label{eq:CRUD}
a(\textup{t}_0)\;=\;\kappa^{\varpi/2}.
\end{equation}
Notice that equations (\ref{eq:SSA}) and (\ref{eq:SSAA}) imply
that
\begin{equation}\label{eq:SSAAA}
|\,\textup{w}_a|\;<\;3\,[1\;+\;\textup{w}_0]^2.
\end{equation}

Unfortunately the current uncertainties in the data for the
quantities H$_0$, w$_0$, and w$_a$ still allow many possibilities.
To give a concrete example, let us adopt ``Hubble units" in which
the currently measured value of H$_0$ is unity. For the sake of
argument we assume that w$_0$ is about $-$ 0.8; this may in fact
be too large: it is compatible with \cite{kn:riess} but is
questionable in view of \cite{kn:tegmark}. According to the
inequality (\ref{eq:SSAAA}), the theory itself predicts that
$|\,\textup{w}_a|$ can be no larger than 12 percent. Any value of
this magnitude is compatible with all observations
\cite{kn:riess}\cite{kn:khoury}\cite{kn:tegmark} [though again the
reader should consult \cite{kn:bass}; see also \cite{kn:bo}].

If we take $|\,\textup{w}_a|$ to be 8 percent, then L is 0.707 in
Hubble units, $\varpi$ is 10, so that we are in the situation
portrayed in the upper panel of Figure 1. The current value of
t$_0$/L is $-$ $\varpi\alpha$ = $-$ 6.16, and that of the scale
function is [by (\ref{eq:CRUD})] about 0.132. If we take
$|\,\textup{w}_a|$ to be 11.6 percent, then L drops to 0.186
Hubble units, $\varpi$ increases to 100, and we are in the lower
panel of Figure 1. The current values of t$_0$/L and of the scale
function are $-$ 18.4 and 0.184 respectively. Examining these
points on the graphs, we see that in both cases the present time
corresponds to a point close to the left side of the base of the
``hill". However, this is not a ``cosmic coincidence": it arises
because we are, for illustrative purposes, deliberately choosing
extreme values of w$_0$ and $|\,\textup{w}_a|$. In fact, of
course, w$_0$ could easily be much closer to $-$1 than the value
we are discussing, and $|\,\textup{w}_a|$ could be much closer to
zero; this is indeed suggested by the latest data analysis
\cite{kn:tegmark}. That would push the present time farther to the
left on the diagrams.

If L is very small in Hubble units, a situation which, as we have
discussed, may arise naturally in Freund-Rubin compactifications
and from the point of view of the dual boundary theory, then
(\ref{eq:QQ}) and (\ref{eq:RR}) give
\begin{equation}\label{eq:CRUDD}
\varpi \;\approx \; {{2}\over
{3\,[1\;+\;\textup{w}_0]}}\;\textup{L}^{-\,2},
\end{equation}
so that the growth of $\varpi$ over-compensates for the smallness
of L: that is, large values of $\varpi$ may be typical from the
holographic point of view.

Our model does not incorporate any kind of matter or radiation
other than $\varphi^-$, so it is pointless to attempt to give a
more detailed numerical account of the observed data in terms of
this model. Broadly speaking, however, the inclusion of matter and
radiation would of course change the shape of the graphs in Figure
1 at very early [and very late] times. In fact, moving towards the
left, the graphs should turn down at a time corresponding to the
observed \cite{kn:riess} onset of acceleration. This will cut off
the long, flat regions of the graphs. [It turns out that the
inclusion of higher modes in equation (\ref{eq:ABRACADABRA2}) also
has this effect; we hope to return to a discussion of this point
elsewhere.]

The inequality (\ref{eq:SSAAA}) means that the theory requires w
to change very slowly if, as is in fact the case, w$_0$ is close
to $-$1. In turn, values of w$_0$ close to $-$1 are suggested by
the fact that the theory favours large values of $\varpi$. That
is, the theory itself requires that observations should continue
to point towards values of w$_0$ close to $-$1 and towards small
values of w$_a$. Contrary to what is often said, then,
\emph{observational findings of that kind} --- see
\cite{kn:tegmark} --- \emph{in no way confirm the idea that the
dark energy corresponds to a [positive] cosmological constant},
since it is now clear that they can be naturally described in this
very different way. As Figure 1 shows [see also Figures 2 and 3
below], the global structure of this temporarily accelerating
universe is very different indeed from that of any de Sitter-like
model. This strongly underlines the fact, stressed in many recent
data analyses
\cite{kn:linder}\cite{kn:alam}\cite{kn:bassett}\cite{kn:melchiorri}\cite{kn:sahni},
that observations showing that w varies slowly and is currently
close to $-$1 are far from proving the existence of a positive
cosmological constant.

We claim, then, that anti-de Sitter physics may be able to account
for the cosmological observations just as well as its much less
well-understood de Sitter counterpart. Of course, one would have
to develop a realistic model containing matter and radiation to
give a detailed account, but it is reasonable to hope that at
least the main virtues of our ``toy" spacetime will survive. This
already represents progress, since there is hope of linking [the
Euclidean version of] anti-de Sitter physics to string theory.
However, as the price to be paid is a Big Crunch, one may wonder
whether this approach is still susceptible to the more general
criticisms of accelerating universes given in, for example,
\cite{kn:susskind}.

A well-known unpleasant feature of de Sitter-based cosmologies is
the ``cosmic coincidence" problem mentioned earlier. This problem
is best formulated as follows. In a spacetime containing nothing
but ordinary matter and radiation together with a positive
cosmological constant, the Universe lasts for an infinite time if
it does not re-collapse before accelerating. In that case, the
graph of the scale function has precisely \emph{one} point of
inflection, and there is an infinite interval of time in which,
granted that we observe acceleration, we might expect to find
ourselves. It is therefore remarkable that in fact we find
ourselves very close \cite{kn:riess} to the point of inflection.
In our cosmology this problem is greatly alleviated. According to
the present model, the cosmic acceleration will soon, by
cosmological standards, come to an end, and therefore our current
location in time is very much less remarkable than it would be in
a de Sitter-like spacetime. This is a welcome alternative to less
palatable solutions of the problem, such as anthropic ones.

More specific criticisms of de Sitter-like cosmologies focus on
their causal structure. There are two main criticisms of this
kind, and it is important that they be kept separate [even though
they have a common mathematical origin]. The first criticism is
that, in de Sitter spacetime, it is possible to find many pairs of
points such that the future null cones of those points never
intersect. Correlations between such events are therefore not
measurable, and it is difficult to reconcile this with the
existence of an S-matrix. Precisely this same observation applies
to the Big Crunch of the traditional ``closed" FRW cosmologies,
for the same reason: the future singularity is spacelike, just as
is the ``infinitely inflated future" of de Sitter spacetime.
However, the physical situation is very different in the case of a
Big Crunch. In de Sitter spacetime, the future null cones fail to
intersect even though they correspond to events on worldlines of
objects which continue to exist for an arbitrarily long period of
time. In a Big Crunch, by contrast, the inability of two observers
to communicate beyond a certain time has a very simple
explanation: they are about to be destroyed, or the Universe is
about to become opaque, and so on. The failure of communication in
such a situation is not a mystery and surely does not in itself
indicate a failure of the S-matrix formalism.

The second criticism is quite different, though it too has its
origin in the spacelike nature of future infinity. In de Sitter
spacetime each inertial observer is surrounded by a horizon,
demarcating the region of spacetime beyond which no event can ever
influence that observer, and this horizon apparently has physical
[thermal] properties despite being observer-dependent. This leads
to many complications \cite{kn:banks} and apparent paradoxes which
it would be best to avoid. Observers in Big Crunch cosmologies
also are [typically] surrounded by such horizons, so it is
important to investigate the parallel argument in our case.

To see what is happening here, let us temporarily allow the
``maximal radius" A of the spatial tori in equation (\ref{eq:L})
to tend to infinity. [We do this only to clarify the subsequent
argument: we continue to insist that cosmic holography probably
requires compact spatial sections, as we discussed earlier.] Now
it is easy to see that the Lorentzian spacetime with metric
(\ref{eq:L}) is conformal to at least part of Minkowski spacetime.
To see which part, define a parameter $\lambda$ by
cos(t$^-$/$\varpi$L) = sech($\lambda$), so that the extent of
conformal time from the Bang until the point of maximum expansion
[t$^-$ = $\lambda$ = 0] is given by
\begin{equation}\label{eq:Q}
\varpi
\textup{L}\int_{-\infty}^{0}\textup{cosh}^{\varpi\;-\;1}(\lambda)d\lambda.
\end{equation}
This diverges precisely in the case of physical interest, that is,
when the spacetime admits a non-zero period of acceleration: for
we saw when discussing equation (\ref{eq:O}) that $\varpi \;>\;1$
is the condition for this.
\begin{figure}[!h]
\centering
\includegraphics[width=0.25\textwidth]{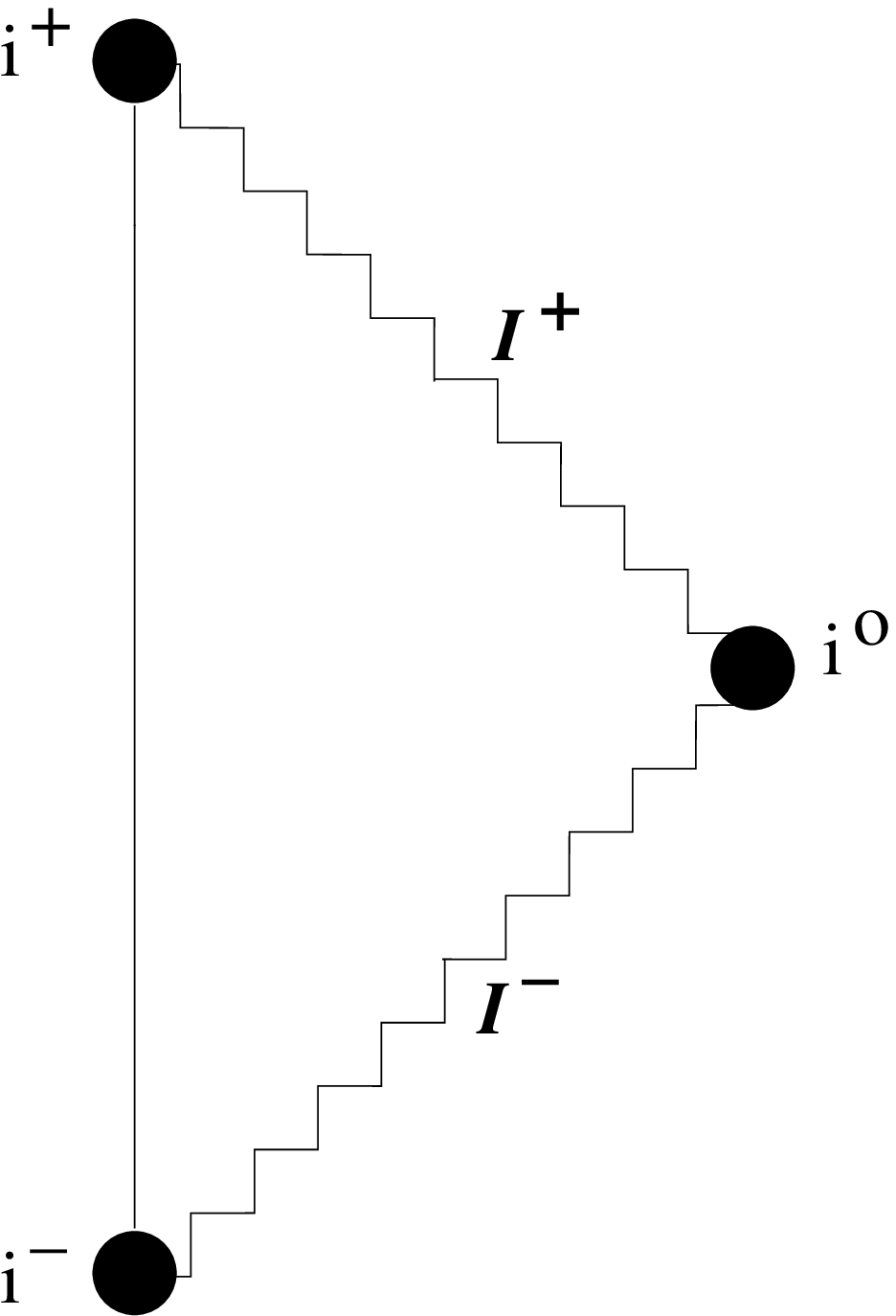}
\caption{Penrose diagram, $\bbr^3$ Spatial Sections}
\end{figure}
Similarly the Crunch is at positive infinity in conformal time if
the Universe accelerates. Thus, in the physically interesting
case, the spacetime is conformal to \emph{all} of Minkowski
spacetime, and so the Penrose diagram is as in Figure 2. [See the
Penrose diagram for Minkowski spacetime, page 123 of
\cite{kn:hawking}.] The jagged lines represent singularities at
future and past null infinity; future and past timelike infinity
are likewise singular, but spatial infinity is of course
non-singular.

Evidently there do \emph{not} exist pairs of points in this
spacetime with non-intersecting future null cones; there is always
some observer who can correlate the two points. Furthermore,
\emph{every} event in this spacetime is visible to a sufficiently
long-lived observer: there is no horizon of the de Sitter kind.
Note that all this holds \emph{provided} that $\varpi\;>\;1$, that
is, provided that the Universe has a period of acceleration. Thus
the objections to de Sitter spacetime discussed above can be
avoided here, and, ironically, it is precisely the presence of a
period of acceleration which allows us to avoid them. The point is
that a Big Crunch need not have a similar conformal structure to
the infinitely inflated future of de Sitter spacetime, and hence
need not have the same objectionable features.
\begin{figure}[!h]
\centering
\includegraphics[width=0.10\textwidth]{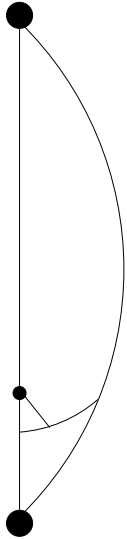}
\caption{Penrose diagram, T$^3$ Spatial Sections}
\end{figure}

If we re-impose a compact [toral] structure for the spatial
sections, the effect is to abolish the null and spatial conformal
infinities. Past and future timelike infinity are then just
points, indicated by the large dots at the top and bottom of
Figure 3. To understand Figure 3, the reader can imagine that the
interior of a spatial torus is an infinite set of nested cubes,
each cube being represented by the two-sphere in which it can be
inscribed. There is a limiting sphere of this kind, and Figure 3
is obtained by deleting all larger two-spheres from Figure 2. The
precise shape of the diagram will of course depend on A, the
maximal ``radius" of the spatial torus. Thus the figure is not
entirely accurate, because more should be deleted in certain
directions, but it conveys the important information here. [This
spacetime, like all spacetimes with flat but topologically
non-trivial spatial sections, is not globally isotropic, so it
cannot be fully represented on a conventional Penrose diagram.]
 Note that careful study of Figure 3 reveals the
fact that the spacetime is in fact \emph{null}, though not of
course timelike,
 geodesically complete.

In Figure 3, as in Figure 2, there are no horizons of the de
Sitter kind. This is an attractive way of avoiding the various
objections which have been raised against de Sitter-like
cosmologies. The curved ``horizontal" line near the bottom of
Figure 3 corresponds to the moment when the Universe became
transparent; the small dot represents the present time; its past
lightcone is also indicated, under the assumption that we are not
\emph{yet} able to see an entire spatial section. This is based on
the pessimistic/conservative assumption that there is no evidence
of topological non-triviality in the present cosmic background
radiation.

This last assumption is in fact somewhat controversial
\cite{kn:cornish}\cite{kn:roukema}; we are taking this
conservative viewpoint here only because we wish to stress that
the compactness of the spatial sections is of fundamental
importance in cosmological holography, \emph{whether or not the
corresponding non-trivial topology is presently observable.} Apart
from the technical advantages mentioned earlier, spatial
compactness will surely reveal itself in the dual theory, for the
following simple reason: the bulk metric does not induce a metric
at infinity, \emph{only a conformal structure}. The question as to
whether the spatial sections of the Universe are too large for
their finiteness to be directly observable is therefore entirely
irrelevant in the dual theory, where the concept of size is
undefined. Indeed, if we can establish a holographic duality of
the kind envisioned by Maldacena and Maoz, it should be capable of
answering all questions regarding the topology of the spatial
sections. Note from Figure 3 that the spatial sections will
\emph{eventually} be completely visible according to our model.

In this section we have discussed an extremely simple example in
which the introduction of matter into AdS\4 produces a spacetime
which naturally reproduces the observed acceleration of the
Universe, and which may in fact improve on de Sitter spacetime in
important ways. Obviously the model is far too simple to be
regarded as the basis of a completely realistic cosmology; and on
the theoretical side, too, much remains to be done. The next step
would be to study in detail the dual field theory at infinity, the
existence of which is the basic hypothesis of holography. Before
we can hope to do that, however, there is a more fundamental issue
which must be settled. We must answer the objection that, since
the Euclidean versions of Maldacena-Maoz cosmologies apparently
reside in the interior of manifolds-with-boundary having
disconnected boundaries [see for example equation (\ref{eq:M})],
these spacetimes actually \emph{contradict} the holographic
principle \cite{kn:yau}. We now turn to this question.

\addtocounter{section}{1}
\section*{4. Are There Really \emph{Two} Boundaries?}

It seems obvious from equation (\ref{eq:M}) that the ``infinity"
of the Euclidean version of our spacetime is a space with two
connected components, one each at t = $\pm\infty$. However,
\emph{this could be a coordinate effect}: recall that it also
``seems obvious" that the Schwarzschild and de Sitter metrics are
singular at the respective horizons if one uses certain
coordinates. Note that care is needed here, because ``infinity" is
made meaningful in these constructions by means of a conformal
transformation which ``makes infinity finite", and the structure
of the resulting finite space is not always obvious.

The following example should make the point clear. Consider a
four-dimensional flat cubic torus with all circumferences equal to
4L and with angular coordinates $\theta$, $\theta_1$, $\theta_2$,
and $\theta_3$; here the angular coordinates all run from $-\pi$
to $+\pi$. The metric is
\begin{equation}\label{eq:STUN}
g^+(\textup{Torus}) \;=\;
{{4\textup{L}^2}\over{\pi^2}}\;[\,d\theta^2 \; +\; d\theta_1^2
\;+\; d\theta_2^2 \;+\; d\theta_3^2\,].
\end{equation}
Now suppose that we conformally re-scale this metric along the
$\theta$ direction [only], in the following way:
\begin{equation}\label{eq:STUNNER}
g^+(2,{{2\textup{L}}\over{\pi}}) \;=\;
{{4\textup{L}^2}\over{\pi^2}}\;[\,1\;-\;{{\theta^2}\over{\pi^2}}\,]^{-2}\;[\,d\theta^2
\; +\; d\theta_1^2 \;+\; d\theta_2^2 \;+\; d\theta_3^2\,];
\end{equation}
the reason for the notation will soon become apparent. The
four-torus is deformed: as we approach the three-torus labelled
$\theta$ = $\pm\pi$ from either side, the three-tori labelled by
$\theta$ are enlarged, the single torus at $\pm\pi$ being
``infinitely large", and also infinitely far from every other
point in the four-torus. Thus the metric is singular at that
three-torus, which means that the metric is really defined on the
space obtained by excising this three-torus from the four-torus.

Going in the reverse direction, presented with a space with the
metric in (\ref{eq:STUNNER}), one can strip away the singular
factor, add in a three-torus at $\theta$ = $\pm\pi$, and obtain
the four-torus with metric (\ref{eq:STUN}) as the ``conformal
compactification" of the space with metric (\ref{eq:STUNNER}). Of
course, T$^4$ is not a conformal compactification in the usual
sense, since it is not a manifold-with-boundary. Nevertheless it
does all that can be expected of a ``conformal compactification",
since it is compact and it allows us to treat infinity as a
``place". This ``place" is a distinguished submanifold instead of
a boundary.

It is of course possible to interpret the space on which
(\ref{eq:STUNNER}) is defined in a more orthodox way. We can do
this by changing the coordinates. Define a new coordinate t$^+$,
with dimensions of length, by
\begin{equation}\label{eq:STUNRATS}
\textup{t}^+\;=\;2\,\textup{L}\,\textup{tanh}^{-1}(\theta/\pi).
\end{equation}
Clearly t$^+$ runs from $-\,\infty$ to $+\,\infty$. Now
transforming the coordinate $\theta$ in equation
(\ref{eq:STUNNER}) accordingly, we find that
$g^+(2,{{2\textup{L}}\over{\pi}})$ is given by
\begin{equation}\label{eq:MONSTER}
g^+(2,{{2\textup{L}}\over{\pi}}) \; = \;(dt^+)^2\; +\;
{{4\textup{L}^2}\over{\pi^2}}\;\textup{cosh}^4({{t^+}\over {2
\textup{L}}})\;[\,d\theta_1^2 \;+\; d\theta_2^2\; +\;
d\theta_3^2\,].
\end{equation}
\emph{But this is nothing but a special case} [$\varpi$ = 2,  A =
2L/$\pi$] of our metric $g^+(\varpi,A)$ given in the general case
by equation (\ref{eq:M}). We repeat: equations (\ref{eq:STUNNER})
and (\ref{eq:MONSTER}) represent \emph{exactly the same metric},
expressed in different coordinates. And yet we agreed that the
``conformal compactification" for (\ref{eq:STUNNER}) was a
four-torus, with infinity corresponding to \emph{one} three-torus
at $\theta$ = $\pm\pi$, while (\ref{eq:MONSTER}) is now seen to be
a special case of a family of manifolds which ``obviously" had
\emph{two} three-tori as the infinity of their conformal
compactifications. What is happening here?

The answer is simple. The basic idea of a Penrose compactification
is that it is often possible to render singular metrics
non-singular by means of a conformal transformation\footnote{In
fact, it is worth noting that this is \emph{always} possible in
the Euclidean case: see \cite{kn:ozeki}.}. The resulting metric is
then defined on a topologically open space. This space can then be
regarded as a subspace of a \emph{compact} space. It is not
generally appreciated, however, that \emph{there is no unique way
of performing this last step}. For example, a space with topology
$\bbr^2$ can be regarded as the interior of a closed
two-dimensional disc; one does this when constructing the
conformal compactification of hyperbolic space H$^2$ [which has
$\bbr^2$ topology]. On the other hand, ordinary stereographic
projection allows us to represent $\bbr^2$ as a subspace of the
sphere S$^2$. Similarly, an open cylinder (0, 1) $\times$ S$^1$
can be represented \emph{either} as a subspace of the compact
closed cylinder [0, 1] $\times$ S$^1$ \emph{or} as a subspace of
the torus T$^2$, by deleting a circle from the latter. This is
precisely the kind of ambiguity we are finding here.

In the analysis of singularities in classical general relativity,
this ambiguity is not important because the conformally related
spacetime is unphysical --- it is a mathematical device. In the
context of holography, however, one has to be more careful,
because there is assumed to be a physical relationship between the
various parts of the compactified space. We must choose our method
of conformally compactifying in the physically most reasonable
way. Clearly, in the cosmological context, that means that we
should do it in such a way that infinity is connected. Let us see
how this works in general.

We shall change the coordinate t$^+$ in equation (\ref{eq:M}) as
follows. Define a constant c$_{\varpi}$ by
\begin{equation}\label{eq:R}
\textup{c}_{\varpi}\; =
\;{{\varpi}\over{\pi}}\int_{0}^{\infty}\textup{sech}^{\varpi}(\zeta)d\zeta
\;.
\end{equation}
Now define a new coordinate $\theta$ by
\begin{equation}\label{eq:RATS}
\textup{c}_{\varpi}\textup{Ld}\theta \;=
\;\pm\textup{sech}^{\varpi}({{t^+}\over {\varpi \textup{L}}})dt^+,
\end{equation}
where the sign is chosen as + when t$^+$ is positive, $-$ when
t$^+$ is negative. The range of $\theta$ is just $-\pi$ to $+\pi$,
corresponding to t$^+$ ranging from $-\infty$ to $+\infty$; that
is, the asymptotic regions are at the finite $\theta$ values
$\pm\pi$, and $\theta$ = 0 corresponds to t$^+$ = 0. Now solve for
t$^+$ in terms of $\theta$ and use this to express
sech$^{\varpi}$(${{\textup{t}^+}\over {\varpi \textup{L}}}$) in
terms of $\theta$. Denote this function by G$_{\varpi}(\theta)$;
then G$_{\varpi}(\theta)$ vanishes at $\pm\pi$, and
$g^+(\varpi,A)$ is given in terms of the coordinate $\theta$ as
\begin{equation}\label{eq:S}
g^+(\varpi,A) \;=\;
\textup{c}^2_{\varpi}\textup{L}^2\;\textup{G}^{-2}_{\varpi}(\theta)\;[\,d\theta^2
\; +\;
({{A}\over{\textup{c}_{\varpi}\,\textup{L}}})^2\,(d\theta_1^2
\;+\; d\theta_2^2 \;+\; d\theta_3^2)].
\end{equation}
This way of expressing the metric is canonical in the sense that
it reveals the fact that this space is \emph{globally} conformally
flat [that is, it has not just the local conformal geometry but
also the \emph{topology} of a flat manifold, a much stronger
property than the mere vanishing of the Weyl tensor]. Here we are
thinking of the ``spatial" sections as cubic tori, as usual. The
number A/(c$_{\varpi}$L) will be discussed below, but for the
moment we shall set it equal to unity. Then (\ref{eq:S}) takes the
particularly simple form
\begin{equation}\label{eq:T}
g^+(\varpi,A)\; =\;
\textup{c}^2_{\varpi}\textup{L}^2\;\textup{G}_{\varpi}(\theta)^{-2}\;[\,d\theta^2
\;+\; d\theta_1^2 \;+\; d\theta_2^2 \;+\; d\theta_3^2].
\end{equation}
The conformal factor diverges at $\theta$ = $\pm\pi$, so this
represents ``infinity" here. But does $\theta$ = $\pm\pi$
represent one region or two? We now know the answer: removing the
conformal factor, we see at once that the structure of the
underlying space is most naturally interpreted as that of a
four-dimensional Euclidean cubic torus. ``Infinity" is just
\emph{one} three-torus at \emph{one} point on the circle
parametrized by $\theta$. Here infinity is represented not by a
boundary component but rather by a special submanifold of the
underlying four-torus. There is no physical reason for insisting
that infinity is a boundary; it is possible to adapt the usual
bulk-infinity correspondence \cite{kn:witten2} to the submanifold
interpretation. The bulk is the open submanifold of T$^4$ obtained
by deleting the T$^3$ at infinity.

In summary, then, the claim that ``infinity is disconnected" for
the space with metric (\ref{eq:M}) is only valid if we insist that
infinity \emph{must} be interpreted mathematically as a boundary.
If we allow a slightly less orthodox interpretation, such as we
propose here, then this is no longer the case. From this new point
of view, the fact that infinity appears to be disconnected in
(\ref{eq:M}) is just due to a bad choice of coordinates. Note that
we are \emph{not} of course claiming that the number of connected
components of the boundary of a manifold-with-boundary can be
changed by means of a change of coordinates: what we are claiming
is that infinity need not be interpreted in terms of
\emph{boundaries} at all. This is in fact a variant of a known
trick in topology --- one is always free to interpret a compact
manifold-with-boundary as a subspace of a compact \emph{manifold},
often called its ``double", and it is often advantageous to do so.

Because our space is globally conformally flat, it makes sense to
speak of the shape [though not of the \emph{size}] of the
underlying four-torus. This shape is of physical interest, for it
can be probed by conformally invariant physical fields, such as
the Yang-Mills fields considered by Maldacena and Maoz
\cite{kn:maoz}. One might suspect that it has some thermodynamic
interpretation, as is usually the case when Euclidean ``time" is
cyclic. Furthermore, if the $\theta$ direction is enormously
larger than the others, then one could object that the bulk on the
two sides of ``infinity" is effectively disconnected for physical
purposes, and then we would be trading one failure of holography
--- two infinities for one bulk
--- for another: ``two" bulks for one infinity. Fortunately
this is a question which can be settled by an appeal to the
observational data.

The shape of the conformal four-torus is determined by the
parameter A/(c$_{\varpi}$L), which in turn is constrained by the
data as follows. Let us assume that the current radius of the
spatial three-torus is K Hubble distances, where K is at least
unity \cite{kn:lahav}, though of course it could be far larger.
According to equation (\ref{eq:CRUD}) we therefore have in Hubble
units
\begin{equation}\label{eq:X}
\textup{A}\;\approx\;\textup{K}\;\kappa^{-\,\varpi/2}.
\end{equation}
If for simplicity we pick $\varpi$ to be the even integer 2n, then
one can show by solving a recursion relation that
\begin{equation}\label{eq:W}
\textup{c}_{2n}\;=\; {{2^{2n}(n!)^2}\over{\pi\;(2n)!}}.
\end{equation}
Thus for the situation portrayed in the upper panel of Figure 1,
we have A $\approx$ 7.59K, c$_{10}$ $\approx$ 1.29, and since L is
0.707 in that case, we see that A/(c$_{\varpi}$L) $\approx$ 8.32K.
In the situation portrayed in the lower panel of Figure 1,
A/(c$_{\varpi}$L) is only a little smaller, A/(c$_{\varpi}$L)
$\approx$ 7.32K. For large values of $\varpi$, Stirling's formula
n! $\approx$
$\sqrt{2\pi\textup{n}}\;\textup{n}^{\textup{n}}\;\textup{e}^{-\;\textup{n}}$
gives c$_{\varpi}$ $\approx$ $\sqrt{\varpi/2\pi}$ and then we find
\begin{equation}\label{eq:Y}
{{\textup{A}}\over{\textup{c}_{\varpi}\textup{L}}}\; \approx\;
\textup{K}\,\sqrt{3\pi(1+\textup{w}_0)}\;\kappa^{({{1}\over{2}}\,-\,[3(1+w_0)(1-\kappa)]^{-1})}.
\end{equation}
This function approaches a minimum value when w$_0$ is taken as
large as possible [compatible with observations] and w$_a$ is
taken as large as possible [compatible with (\ref{eq:SSAAA})]. If,
as usual, we take w$_0$ to be $-$0.8, then this value is about
3.86K, so A/(c$_{\varpi}$L) cannot be less than this. Thus the
underlying conformal four-torus is far from being elongated in the
$\theta$ direction: it is in fact shorter in that direction than
in the other three directions, by a factor between about 4 and 8,
even in the most optimistic case in which the current radius of
the spatial torus is about one Hubble distance. Thus we certainly
have one bulk for one infinity.

We have seen that it is not possible to \emph{prove} that equation
(\ref{eq:M}) represents a space which \emph{must} be regarded as
the interior of a manifold-with-boundary having two boundary
components. For we have also represented the space as an open
submanifold of a compact manifold with no boundary at all;
infinity then corresponds to a \emph{connected} submanifold of
that compact manifold. This is not to say that the question as to
whether infinity is connected is meaningless. Rather, it means
that the question is one which has to be settled by physics, not
merely by inspecting the form of the metric tensor. [The same
remark applies to the structure of the ``spatial" sections: we
have assumed for simplicity that they are tori rather than one of
the other nine compact flat three-manifolds, but this is not
imposed on us by the form of the metric
--- it is a physical question to be settled by physical
arguments.] We shall now address this question.

We saw that the requirements that the Lorentzian version of our
spacetime should accelerate, and that the Euclidean version should
be topologically non-trivial, led us directly to the Euclidean
axion as the most natural candidate for the matter content of our
model. Now combining equations (\ref{eq:J}), (\ref{eq:K}), and
(\ref{eq:RATS}), we find that $\theta$ and $\varphi^+$ are related
by
\begin{equation}\label{eq:ROCKS}
\textup{c}_{\varpi}\textup{d}\theta \;= \; \sqrt{4\pi\varpi}\;
                \textup{cos}^{\varpi\;-\;1}(\sqrt{{{4\pi}\over
                {\varpi}}}\,\varphi^+)\,\textup{d}\varphi^+;
\end{equation}
for example, if $\varpi$ = 2, the relation is just $\theta$ =
$\pi\;$sin($\sqrt{2\pi}\;\varphi^+$). This means that we can
interpret $\varphi^+$ as a \emph{coordinate} which could be used
to replace $\theta$. As $\theta$ ranges from $-\pi$ to $+\pi$,
$\varphi^+$ ranges from
($-\sqrt{{{\varpi}\over{4\pi}}}\times{{\pi}\over{2}}$) to
($+\sqrt{{{\varpi}\over{4\pi}}}\times{{\pi}\over{2}}$). These are
the values of $\varphi^+$ corresponding to the two asymptotic
regions. The question as to whether there are really two
boundaries [that is, whether $\theta$ is really an angular
variable] should therefore be formulated as: are there physical
reasons for interpreting $\varphi^+$ as an angular coordinate?

Since $\varphi^+$ is an axion, the answer may seem obvious. But
there is an interesting subtlety here: let us ask again: do the
two extreme values of $\varphi^+$ correspond to two different
physical situations?

If one begins with equation (\ref{eq:E}), which can be interpreted
as the result of introducing a certain kind of matter [with
potential $\textup{V}^+_{\textup{Axion}}$] into Euclidean AdS\4,
then it is indeed natural to claim that the two situations are
identical. For the function $\textup{V}^+_{\textup{Axion}}$ is
\emph{periodic with period} $\sqrt{{{\varpi}\over{4\pi}}}\times
\pi$, so the $\varphi^+$ values
($-\sqrt{{{\varpi}\over{4\pi}}}\times{{\pi}\over{2}}$) and
($+\sqrt{{{\varpi}\over{4\pi}}}\times{{\pi}\over{2}}$) should not
be distinguished. From this point of view, then, it is clear that
the Euclidean metric given by (\ref{eq:M}) is defined on the open
submanifold of the four-torus given by deleting a three-torus. The
conformal compactification is the full four-torus, T$^4$, with its
metric visible in (\ref{eq:S}) after removing the conformal
factor; \emph{here all of the coordinates are angular
coordinates}. More generally, since we do not know the topology of
the spatial sections of our universe, the topology of the
compactification is S$^1\,\times\,$(T$^3$/F), where F can be
\emph{any} of the entries in the known list of finite groups
\cite{kn:wolf} which can be holonomy groups of compact
three-dimensional flat Riemannian manifolds.

On the other hand, one may feel that the superpotential
W$^+(\varphi^+)$ is more fundamental than the potential derived
from it. This is in fact implicit in the procedure we have
followed in this paper, where we are taking the Euclidean point of
view to be fundamental, and in which we began by postulating a
Euclidean superpotential given by equation (\ref{eq:C}). Notice
that while the superpotential always occurs quadratically in the
formula for the potential, it does not always do so when couplings
to spinors are considered \cite{kn:town}; therefore we cannot
ignore its sign. In view of this, it seems now that the two ends
of the range for $\varphi^+$ are \emph{not} identical, since
\begin{equation}\label{eq:V}
\textup{W}^+(-\sqrt{{{\varpi}\over{4\pi}}}\times{{\pi}\over{2}})\;=\;-\;
\textup{W}^+(+\sqrt{{{\varpi}\over{4\pi}}}\times{{\pi}\over{2}}).
\end{equation}
Thus the superpotential detects the difference even though the
potential cannot. \emph{Notice that the periodic identification we
seek is not an automatic consequence of the fact that we began
with a [Euclidean] axion}. Indeed, it begins to seem that an
axionic superpotential may actually forbid a cyclic interpretation
of Euclidean `` conformal time".

In fact, however, this problem arises from taking a too simplistic
view of the process of ``identifying the ends". We have stressed
that our metric allows for the possibility that the topology of
the ``spatial" sections of the Euclidean version of the space
[which is identical to the topology of the true spatial sections
in the Lorentzian version] could be any one of the 10 possible
topologies \cite{kn:wolf} for flat compact three-manifolds. One
can think of this in the following way: when constructing a torus,
one identifies opposite sides of a parallelopiped, but it is also
possible to do this after performing ``twists" of varying degrees
of complexity. This can of course also be done in \emph{four}
dimensions. There are 75 distinct possible topologies for compact
four-dimensional flat manifolds, most of which are non-orientable:
there are 48 non-orientable spaces and 27 orientable ones. Thus,
in performing the identification suggested so strongly by
equations (\ref{eq:S}) and (\ref{eq:T}), we are free to perform
such a twist.

In particular, we can define a compact flat four-dimensional
manifold as follows. Let a$_{\mu}$, where $\mu$ = 1 through 4, be
an orthogonal basis for $\bbr^4$, where we take a$_1$ to be of
unit length, while a$_2$, a$_3$, and a$_4$ are of length
A/c$_{\varpi}$L. For any real $\beta$ let $\beta\,$T$_{\mu}$
denote the isometry of $\bbr^4$ [with its standard metric] defined
by translation by $\beta\,$a$_{\mu}$. Let B be the isometry of
$\bbr^4$ defined by the linear extension of
\begin{equation}\label{eq:ZA}
\textup{B}\; : \;a_1\; \rightarrow \;a_1,\;\;\;\;\;\; \textup{B} \;: \;a_2
\;\rightarrow\; -\,a_2,\;\;\;\;\;\; \textup{B} \;:\; a_3 \;\rightarrow\; -\,a_3,
 \;\;\;\;\;\;\textup{B}\; :\;
a_4 \;\rightarrow \;-\,a_4,
\end{equation}
and consider the $\bbr^4$ isometry defined by $\alpha$ = (B,
$\frac{1}{2}\,$T$_1$), meaning B followed by
${{1}\over{2}}\,$T$_1$. Clearly $\alpha^2$ = T$_1$, and so if
$\alpha\,\zeta$ = $\zeta$ for some $\zeta$ in $\bbr^4$ then
$\zeta$ is also a fixed point of T$_1$, which is impossible. In
fact the non-abelian infinite group $\Gamma$ generated by $\alpha$
and the T$_{\mu}$ acts properly discontinuously and freely on
$\bbr^4$, so $\bbr^4/\Gamma$ is a smooth manifold. To be more
precise, $\Gamma$ is a group with no element of finite order
[other than the identity] and with a maximal free abelian subgroup
of index 2 and rank 4 [generated by the T$_{\mu}$]. By the
relevant version of the Bieberbach theorems [\cite{kn:wolf}, page
105] it follows that $\bbr^4/\Gamma$ is a compact flat
four-dimensional manifold. The metric is the one visible in
equation (\ref{eq:S}) after the removal of the conformal factor.
As a Riemannian manifold, $\bbr^4/\Gamma$ can be expressed  as
T$^4$/$\bbz_2$, where T$^4$ is the rectangular oblate torus with
aspect ratio given by A/(c$_{\varpi}$L) as discussed earlier, and
where $\bbz_2$ is the linear holonomy group of this space. [This
expression is not unique, but this will not matter here.] This
allows us to think of the manifold as being like a torus which has
been cut through and then glued together after reflecting one end.
In fact we can regard KB$^4$ = $\bbr^4/\Gamma$ as a
four-dimensional version of a Klein bottle.

If an orthonormal basis is parallel transported around this space
along the a$_1$ direction, it will return to the initial point
with orientation reversed in the way defined by B, because B
essentially generates the $\bbz_2$ holonomy group of KB$^4$. The
axion field $\varphi^+$ is able to detect this reversal. The
upshot is that the superpotential W$^+(\varphi^+)$ is no longer
able to distinguish the ``two" infinities, and once again we have
a physical justification for regarding the ``two" as one location,
artificially split by a choice of coordinates. Notice that while
KB$^4$ = T$^4$/$\bbz_2$ is non-orientable, the reversals of
orientation occur only in the ``time" direction: the spatial
sections [and the infinity submanifold] themselves are still
orientable, being in fact copies of the three-torus T$^3$.
\emph{Both} bulk and infinity are orientable separately --- only
the combined space is non-orientable. Hence there is no conflict
with the putative holographic duality, and no complications
arising in connection with formulating spinors on non-orientable
manifolds.

Thus we see that, whether one takes the potential or the
superpotential to be fundamental, it is possible to identify the
``two" boundary components which are \emph{apparently} required by
Maldacena-Maoz cosmologies, that is, by the Euclidean formulation
of a Bang/Crunch cosmology. If, following Witten and Yau
\cite{kn:yau}, one regards a double boundary as a contradiction of
the holographic principle, then surely this is the simplest
possible resolution of this problem.

If we accept that this is the correct response to the ``double
boundary problem", then it is natural to ask what [if anything]
this identification implies for the Lorentzian versions of these
spacetimes. The obvious assumption is that it implies that [in a
sense to be clarified] \emph{conformal} time is cyclic in the
Lorentzian case also. However, it is not at all clear that this is
\emph{necessary}: after all, the essence of the Maldacena-Maoz
proposal is that holography works in the Euclidean domain, not in
the Lorentzian: see Figure 3 . It therefore seems quite plausible
to us that the Lorentzian Bang and Crunch need \emph{not} be
identified. However, let us briefly explore the alternative
possibility. Once again, if we are to argue that the two
components of Lorentzian conformal infinity should be identified,
this must be done on the basis of a physical argument and not
because of a formal analogy.

In the particular case of the spacetime we have been discussing
here, there actually \emph{is} such an argument: it runs as
follows. Equations (\ref{eq:K}) and (\ref{eq:U}) can be written as
\begin{equation}\label{eq:KLOWN}
\textup{cos}(\sqrt{{{4\pi}\over{\varpi}}}\;\varphi^+)\;=\;\textup{sech}({{t^+}\over{\varpi\textup{L}}}),
\end{equation}
\begin{equation}\label{eq:UDDER}
\textup{cos}({{t^-}\over{\varpi\textup{L}}})\;=\;\textup{sech}(\sqrt{{{4\pi}\over{\varpi}}}\;\varphi^-).
\end{equation}
This clearly establishes a natural one-to-one map between the
Euclidean and Lorentzian versions of this spacetime: leaving aside
the constant factors, we can map a given t$^+$ slice of the
Euclidean space to the Lorentzian slice which has t$^-$ given by
the value of $\varphi^+$ corresponding to t$^+$. In that sense,
the Euclidean axion field $\varphi^+$ \emph{is actually Lorentzian
time}. Similarly, of course, the Lorentzian axion \emph{is}
Euclidean ``time". But we have argued that $\varphi^+$ must be
considered to be essentially periodic, meaning that the Euclidean
conformal ``time" coordinate $\theta$ is an angular variable on
the underlying manifold of the Euclidean space. Hence we conclude
that, in this spacetime, the timelike direction is periodic
\emph{after} infinity has been ``made finite": that is, Lorentzian
conformal time is periodic. There are no closed timelike
worldlines in the spacetime itself, but there \emph{are} such
worldlines in the Penrose diagram; in other words, the large dots
at the top and bottom of Figure 3 must be identified. This is the
physical interpretation of the ``twist" [reflection] that occurs
as the four-dimensional Klein bottle KB$^4$ is circumnavigated:
the twist essentially ``re-sets the Lorentzian clock".

The status of closed timelike worldlines in fundamental theories
such as string/M theory is a subject of much current interest; see
\cite{kn:gott} for an extensive list of references to early work
and \cite{kn:johnson}\cite{kn:gutowski} for more recent
references. In the spacetime we are considering here, such
worldlines appear only very indirectly, in the spacetime with the
conformally transformed metric: we stress that there are no closed
timelike worldlines in the physical spacetime itself. However, the
closed timelike worldlines in the conformally related spacetime
may not be completely hidden, since the ``unphysical" spacetime
can be probed by conformally invariant fields such as gauge fields
or curvature coupled scalars. It is therefore reassuring that
there is no evidence in these works suggesting that string/M
theory forbids such worldlines entirely. It would be of great
interest to see how the discussions in \cite{kn:johnson} and
\cite{kn:gutowski} can be adapted to the case where the Crunch is
just the Bang approached from the ``other side".

The identification of the Crunch with the Bang leads to no
difficulties in spacetimes as simple as the one considered here,
but it could be otherwise in a more realistic spacetime model: for
thermodynamic reasons one supposes that conditions in the
contracting phase of a Bang/Crunch cosmology will be quite
different to those in the expanding phase. In the Euclidean
version of the conformally related space, this could mean that
there will be a sudden jump in the geometry as one passes through
$\theta$ = $\pm\pi$. This is exactly the situation for a Euclidean
version of an ``asymmetric brane" spacetime --- see for example
\cite{kn:carter}. Using these techniques one can try to account
for the jump by [at least formally] assigning a stress-energy
tensor to a ``brane" at $\theta\ = \,\pm\pi$ in the conformally
related space. It remains to be seen whether suitable asymmetric
brane-like junction conditions can be formulated in such a way as
to produce a more realistic cosmology.

The identification of the Crunch with the Bang is also an
interesting move from the point of view of the ``cyclic" cosmology
\cite{kn:turok2}. It is not known definitely whether this
cosmology ``cycles" \emph{eternally}; but if it does, then it is
natural to ask whether the successive ``lifetimes" of the cosmos
are really distinct. A cyclic conformal time coordinate would of
course be perfectly natural in such a case. This would be
particularly interesting if it can be established, as suggested in
\cite{kn:turok3}, that string or M theory allows an orderly
evolution through cosmological singularities.

To summarize briefly: AdS$_4$ cosmology leads naturally to an
apparently disconnected boundary for the Euclidean version of the
spacetime. Holography seems to require that the two boundary
components should be identified, an option which is always open in
the cosmological case since the boundary components must have the
same topology. In this section we have given a very simple example
in which this identification is actually suggested by the physics
of the matter content of the spacetime. It turns out that the
identification has two surprising properties: it involves a
non-trivial ``twist" as the identification is performed, and it
may possibly necessitate a similar identification in the
Lorentzian version. The main point is that, contrary to
appearances, these spacetimes are \emph{not} fundamentally
incompatible with the holographic principle. The ``double"
boundary is merely a signal of unexpected complexity in the
topology of the conformal compactification.

 \addtocounter{section}{1}
\section*{5. Conclusion}
The principal lessons of this investigation can be stated very
briefly as follows.

First, there is no evidence, theoretical or [of course]
observational, that the current cosmic acceleration is permanent.
As soon as this is granted, however, one should begin to question
whether the Universe is now or ever has been in a de Sitter-like
state. An alternative is that it could be in an ``anti-de
Sitter-like" state.

Second, however, the result of adding matter to AdS is \emph{not}
a spacetime that resembles AdS: because of the singularity
theorems, a cosmological AdS spacetime will typically have both a
Bang and a Crunch. Thus the spacetimes considered by Maldacena and
Maoz \cite{kn:maoz} are in this sense generic.

Third, the Euclidean versions of the spacetimes so obtained will
present the appearance of having a disconnected conformal
infinity. As is so often the case in curved spacetimes, however,
this is a situation in which coordinates can be very deceptive.
The question as to whether infinity is disconnected \emph{cannot}
be settled by inspecting the form taken by the metric in some
coordinate system.

Fourth, a disconnected infinity can be avoided --- as holography
suggests that it should be --- by accepting that the conformal
compactification may have a more complex topology than appears at
first glance. The degree of complexity should be determined by the
physical nature of the matter content of the spacetime.

Naturally, having clarified the basic way in which holography may
work in these cosmologies, we have yet much to do. One major task
is to determine the precise nature of the field theory at infinity
which corresponds to the axion in the Euclidean bulk. Another is
to understand the role of the topology of the spatial sections of
the Universe: as we have emphasised, this topology is of basic
importance in cosmic holography whether or not it can be observed
at the present time. In particular it is curious that \emph{both}
the spatial sections and the conformal compactification of the
whole [Euclidean] spacetime have the topology of compact flat
manifolds. These spaces are known to have very remarkable
geometric properties --- for example \cite{kn:lawson}, an
initially arbitrary metric on them has to be \emph{exactly} flat
if its scalar curvature vanishes. We intend to discuss the
physical consequences elsewhere.

\addtocounter{section}{1}
\section*{Acknowledgements}
The author is grateful for the kind hospitality and stimulating
environment of the High Energy Section of the Abdus Salam
International Centre for Theoretical Physics, where this work was
done. He also wishes to thank Wanmei for the diagrams and for
making our stay in Italy so thoroughly productive and enjoyable.


\begin{thebibliography}{18}

\bibitem{kn:carroll}
Sean M. Carroll, Why is the Universe Accelerating?, in Measuring
and Modeling the Universe, Carnegie Observatories Astrophysics
Series Vol. 2, ed. W. L. Freedman, \x astro-ph/0310342
\bibitem{kn:riess}
Adam G. Riess et al, Type Ia Supernova Discoveries at z $>$ 1 From
the Hubble Space Telescope: Evidence for Past Deceleration and
Constraints on Dark Energy Evolution, Astrophys.J. 607 (2004) 665,
\x astro-ph/0402512
\bibitem{kn:witten1}
Edward Witten, Quantum Gravity In De Sitter Space, \x
hep-th/0106109
\bibitem{kn:susskind}
Simeon Hellerman, Nemanja Kaloper, Leonard Susskind, String Theory
and Quintessence, JHEP 0106 (2001) 003, \x hep-th/0104180
\bibitem{kn:fischler}
W.Fischler, A.Kashani-Poor, R.McNees, S.Paban, The Acceleration of
the Universe, a Challenge for String Theory, JHEP 0107 (2001) 003,
\x hep-th/0104181
\bibitem{kn:sim}
Brett McInnes, Exploring the Similarities of the dS/CFT and
AdS/CFT Correspondences, Nucl.Phys. B627 (2002) 311, \x
hep-th/0110062
\bibitem{kn:strominger}
A. Strominger, The dS/CFT Correspondence, JHEP 0110 (2001) 034, \x
hep-th/0106099
\bibitem{kn:klemm1}
Sergio Cacciatori, Dietmar Klemm, The Asymptotic Dynamics of de
Sitter Gravity in three Dimensions, Class.Quant.Grav. 19 (2002)
579, \x hep-th/0110031
\bibitem{kn:minic}
V. Balasubramanian, J. de Boer, D. Minic, Exploring de Sitter
Space and Holography, Class.Quant.Grav. 19 (2002) 5655-5700;
Annals Phys. 303 (2003) 59-116, \x hep-th/0207245
\bibitem{kn:goheer} Naureen Goheer, Matthew Kleban, Leonard
Susskind, The Trouble with de Sitter Space, JHEP 0307 (2003) 056,
\x hep-th/0212209
\bibitem{kn:mcinnesschwarz}
Brett McInnes, De Sitter and Schwarzschild-De Sitter According to
Schwarzschild and De Sitter, JHEP 0309 (2003) 009, \x
hep-th/0308022
\bibitem{kn:silver}
Mohsen Alishahiha, Andreas Karch, Eva Silverstein, David Tong, The
dS/dS Correspondence, \x hep-th/0407125
\bibitem{kn:klemm2}
Dietmar Klemm, Luciano Vanzo, Aspects of Quantum Gravity in de
Sitter Spaces, \x hep-th/0407255
\bibitem{kn:cardenas}
Rolando Cardenas, Tame Gonzalez, Yoelsy Leiva, Osmel Martin,
Israel Quiros, A model of the Universe including Dark Energy
accounted for by both a Quintessence Field and a (negative)
Cosmological Constant, Phys.Rev. D67 (2003) 083501, \x
astro-ph/0206315
\bibitem{kn:decay}
Ujjaini Alam, Varun Sahni, A. A. Starobinsky, Can dark energy be
decaying?, JCAP 0304 (2003) 002, \x astro-ph/0302302
\bibitem{kn:cvetic}
Mirjam Cvetic, Shin'ichi Nojiri, Sergei D. Odintsov, Cosmological
anti-deSitter space-times and time-dependent AdS/CFT
correspondence, Phys.Rev. D69 (2004) 023513, \x hep-th/0306031
\bibitem{kn:hawking}
S. W. Hawking, G. F. R. Ellis, The Large Scale Structure of
Space-Time, Cambridge University Press, 1973.
\bibitem{kn:leblond}
Frederic Leblond, Mirage resolution of cosmological singularities,
\x hep-th/0403221
\bibitem{kn:caldwell}
R.R. Caldwell, A Phantom Menace? Cosmological consequences of a
dark energy component with super-negative equation of state,
Phys.Lett. B545 (2002) 23, \x astro-ph/9908168
\bibitem{kn:linde}
Renata Kallosh, Jan Kratochvil, Andrei Linde, Eric V. Linder,
Marina Shmakova, Observational Bounds on Cosmic Doomsday, JCAP
0310 (2003) 015, \x astro-ph/0307185
\bibitem{kn:maldacena}
Juan M. Maldacena, TASI 2003 Lectures on AdS/CFT, \x
hep-th/0309246
\bibitem{kn:horowitz}
Thomas Hertog, Gary T. Horowitz, Towards a Big Crunch Dual, JHEP
0407 (2004) 073, \x hep-th/0406134
\bibitem{kn:maoz}
Juan Maldacena, Liat Maoz, Wormholes in AdS, JHEP 0402 (2004) 053,
\x hep-th/0401024
\bibitem{kn:yau}
Edward Witten, S.-T. Yau, Connectedness Of The Boundary In The
AdS/CFT Correspondence, Adv.Theor.Math.Phys. 3 (1999) 1635, \x
hep-th/9910245
\bibitem{kn:mcinnes}
Brett McInnes, Quintessential Maldacena-Maoz Cosmologies, JHEP
0404 (2004) 036, \x hep-th/0403104
\bibitem{kn:maoz2}
L. Maoz, talk at Strings 2004,
http://strings04.lpthe.jussieu.fr/talks/Maoz.pdf
\bibitem{kn:balasub}
Vijay Balasubramanian, Asad Naqvi, Joan Simon, A Multi-Boundary
AdS Orbifold and DLCQ Holography: A universal holographic
description of extremal black hole horizons, JHEP 0408 (2004) 023,
 \x hep-th/0311237
\bibitem{kn:gukov}
Sergei Gukov, Emil Martinec, Gregory Moore, Andrew Strominger,
Chern-Simons Gauge Theory and the AdS(3)/CFT(2) Correspondence, \x
hep-th/0403225
\bibitem{kn:geroch}
R.P. Geroch, Topology in general relativity. J. Mathematical Phys.
8 (1967) 782
\bibitem{kn:tipler}
F.J. Tipler, Topology change in Kaluza-Klein and superstring
theories, Phys. Lett. B 165 (1985) 67
\bibitem{kn:ratra}
P. J. E. Peebles, Bharat Ratra, The Cosmological Constant and Dark
Energy, Rev.Mod.Phys. 75 (2003) 559, \x astro-ph/0207347
\bibitem{kn:GS}
S.B.Giddings, A. Strominger, Axion-Induced Topology Change in
Quantum Gravity and String Theory, Nucl.Phys. B306 (1988) 890
\bibitem{kn:wald}
Gerard Jungman, Robert M Wald, Nonexistence Theorems for
Asymptotically Euclidean Einstein-Matter Solutions, Phys. Rev. D40
(1989) 2615
\bibitem{kn:coule}
D.H. Coule, No Wormholes With Real Minimally Coupled Scalar
Fields, Phys.Rev.D55 (1997) 6606
\bibitem{kn:choi}
Kiwoon Choi, String or M theory axion as a quintessence, Phys.Rev.
D62 (2000) 043509, \x hep-ph/9902292
\bibitem{kn:kim}
Jihn E. Kim, Hans Peter Nilles, A Quintessential Axion, Phys.Lett.
B553 (2003) 1, \x hep-ph/0210402
\bibitem{kn:gonzalez}
Pedro F. Gonzalez-Diaz, Axion Phantom Energy, Phys.Rev. D69 (2004)
063522, \x hep-th/0401082
\bibitem{kn:witten2}
Edward Witten, Anti De Sitter Space And Holography,
Adv.Theor.Math.Phys. 2 (1998) 253, arXiv:hep-th/9802150
\bibitem{kn:orbifold}
Brett McInnes, Orbifold Physics and de Sitter Spacetime, Nucl Phys
B 692(2004)270, \x hep-th/0311055
\bibitem{kn:brand}
Scott Watson, Robert Brandenberger, Linear Perturbations in Brane
Gas Cosmology, JHEP 0403 (2004) 045, \x hep-th/0312097
\bibitem{kn:watson}
Thorsten Battefeld, Scott Watson, Effective Field Theory Approach
to String Gas Cosmology, JCAP 0406 (2004) 001, \x hep-th/0403075
\bibitem{kn:paban}
W. Fischler, S. Paban, M. Zanic, The Energy Density of ``Wound"
Fields in a Toroidal Universe, \x astro-ph/0407349
\bibitem{kn:lawson}
H. Blaine Lawson and Marie-Louise Michelsohn, Spin Geometry,
Princeton University Press, 1990
\bibitem{kn:wolf}
J.A. Wolf, Spaces of Constant Curvature, Fifth Edition, Publish or
Perish Press, 1984
\bibitem{kn:olive}
Seokcheon Lee, Keith A. Olive, Maxim Pospelov, Quintessence Models
and the Cosmological Evolution of alpha, \x astro-ph/0406039
\bibitem{kn:cline}
J.J. Blanco-Pillado, C.P. Burgess, J.M. Cline, C. Escoda, M.
Gomez-Reino, R. Kallosh, A. Linde, F. Quevedo, Racetrack
Inflation, \x hep-th/0406230
\bibitem{kn:townsend}
P.K. Townsend, Quintessence from M-Theory, JHEP 0111 (2001) 042,
\x hep-th/0110072
\bibitem{kn:ish1}
Ishwaree P. Neupane, Accelerating Cosmologies from Exponential
Potentials, Class. Quant. Grav. 21 (2004) 4383, \x hep-th/0311071
\bibitem{kn:kehagias}
A. Kehagias, G. Kofinas, Cosmology with exponential potentials,
Class.Quant.Grav. 21 (2004) 3871, \x gr-qc/0402059
\bibitem{kn:jarv}
Laur Jarv, Thomas Mohaupt, Frank Saueressig, Quintessence
Cosmologies with a Double Exponential Potential, JCAP 0408 (2004)
016, \x hep-th/0403063
\bibitem{kn:erosh}
E. Babichev, V. Dokuchaev, Y. Eroshenko, Dark energy cosmology
with generalized linear equation of state, \x astro-ph/0407190
\bibitem{kn:turok2}
Paul J. Steinhardt, Neil Turok, The Cyclic Model Simplified, \x
astro-ph/0404480
\bibitem{kn:turok1}
Joel K. Erickson, Daniel H. Wesley, Paul J. Steinhardt, Neil
Turok, Kasner and Mixmaster behavior in universes with equation of
state w $\ge$ 1, Phys.Rev. D69 (2004) 063514, \x hep-th/0312009
\bibitem{kn:bobby}
Bobby S Acharya, Observations on the Space of Four Dimensional
String and $M$ theory Vacua, \x hep-th/0406228
\bibitem{kn:denef}
B.S. Acharya, F. Denef, C. Hofman, N. Lambert, Freund-Rubin
Revisited, \x hep-th/0308046
\bibitem{kn:khoury}
Sheng Wang, Justin Khoury, Zoltan Haiman, Morgan May, Constraining
the Evolution of Dark Energy with a Combination of Galaxy Cluster
Observables, \x astro-ph/0406331
\bibitem{kn:bass}
Bruce A. Bassett, Pier Stefano Corasaniti, Martin Kunz, The
essence of quintessence and the cost of compression, \x
astro-ph/0407364
\bibitem{kn:linder}
Eric V. Linder, Reconstructing and Deconstructing Dark Energy, \x
astro-ph/0406189
\bibitem{kn:tegmark}
U. Seljak et al, Cosmological parameter analysis including SDSS
Ly-alpha forest and galaxy bias: constraints on the primordial
spectrum of fluctuations, neutrino mass, and dark energy, \x
astro-ph/0407372
\bibitem{kn:bo}
Bo Feng, Xiulian Wang, Xinmin Zhang, Dark Energy Constraints from
the Cosmic Age and Supernova, \x astro-ph/0404224
\bibitem{kn:alam}
Ujjaini Alam, Varun Sahni, A. A. Starobinsky, The case for
dynamical dark energy revisited, JCAP 06 (2004) 08, \x
astro-ph/0403687
\bibitem{kn:bassett}
P.S. Corasaniti, M. Kunz, D. Parkinson, E.J. Copeland, B.A.
Bassett, The foundations of observing dark energy dynamics with
the Wilkinson Microwave Anisotropy Probe, \x astro-ph/0406608
\bibitem{kn:melchiorri}
Alessandro Melchiorri, New Constraints on Dark Energy, \x
astro-ph/0406652
\bibitem{kn:sahni}
Ujjaini Alam, Varun Sahni, Tarun Deep Saini, A. A. Starobinsky,
Rejoinder to "No Evidence of Dark Energy Metamorphosis", \x
astro-ph/0406672
\bibitem{kn:banks}
T. Banks, Some Thoughts on the Quantum Theory of de Sitter Space,
\x astro-ph/0305037
\bibitem{kn:cornish}
Neil J. Cornish, David N. Spergel, Glenn D. Starkman, Eiichiro
Komatsu, Constraining the Topology of the Universe, Phys.Rev.Lett.
92 (2004) 201302, \x astro-ph/0310233
\bibitem{kn:roukema}
Boudewijn F. Roukema, Bartosz Lew, Magdalena Cechowska, Andrzej
Marecki, Stanislaw Bajtlik, A Hint of Poincar\'e Dodecahedral
Topology in the WMAP First Year Sky Map, Astron.Astrophys. 423
(2004) 821, \x astro-ph/0402608
\bibitem{kn:ozeki}
K. Nomizu, H. Ozeki, The existence of complete Riemannian metrics,
Proc. Amer. Math. Soc. 12 (1961) 889
\bibitem{kn:lahav}
Ofer Lahav, Andrew R Liddle, The Cosmological Parameters, In "The
Review of Particle Physics", S. Eidelman et al, Physics Letters
B592 (2004) 1, \x astro-ph/0406681
\bibitem{kn:town}
Kostas Skenderis, Paul K. Townsend, Gravitational Stability and
Renormalization-Group Flow, Phys.Lett. B468 (1999) 46, \x
hep-th/9909070
\bibitem{kn:gott}
J. Richard Gott, III, Li-Xin Li, Can the Universe Create Itself?,
Phys.Rev. D58 (1998) 023501, \x astro-ph/9712344
\bibitem{kn:johnson}
Clifford V. Johnson, Harald G. Svendsen, An Exact String Theory
Model of Closed Time-Like Curves and Cosmological Singularities,
\x hep-th/0405141
\bibitem{kn:gutowski}
Jerome P. Gauntlett, Jan B. Gutowski, Nemani V. Suryanarayana, A
deformation of AdS$_5 \times$ S$^5$, \x hep-th/0406188
\bibitem{kn:carter}
Richard A. Battye, Brandon Carter, Andrew Mennim, Jean-Philippe
Uzan, Einstein equations for an asymmetric brane-world, Phys.Rev.
D64 (2001) 124007, \x hep-th/0105091
\bibitem{kn:turok3}
Neil Turok, Malcolm Perry, Paul J. Steinhardt, M Theory Model of a
Big Crunch/Big Bang Transition, \x hep-th/0408083














\end{thebibliography}
\end{document}